\documentclass[preprint,amsmath,amsssymb,aps,prl,groupadress,superscriptaddress,citeautoscript]{revtex4-1}
\usepackage[utf8]{inputenc}
\usepackage{mhchem}
\usepackage{dcolumn}
\usepackage[dvipdfmx]{graphicx}
\usepackage{bm}
\usepackage{color}
\setcitestyle{super}

\date{\today}

\begin{document}
\title{Observation of chiral domain walls in an octupole-ordered antiferromagnet}

\author{Moeta Tsukamoto}
\email{moeta.tsukamoto@phys.s.u-tokyo.ac.jp}
\affiliation{Department of Physics, The University of Tokyo, Bunkyo, Tokyo 113-0033, Japan}
\author{Zhewen Xu}
\affiliation{Department of Physics, ETH Zurich, Otto-Stern-Weg 1, 8093 Zurich, Switzerland}
\author{Tomoya Higo}
\affiliation{Department of Physics, The University of Tokyo, Bunkyo, Tokyo 113-0033, Japan}
\author{Kouta Kondou}
\affiliation{RIKEN, Center for Emergent Matter Science (CEMS), Saitama 351-0198, Japan}
\author{Kento Sasaki}
\affiliation{Department of Physics, The University of Tokyo, Bunkyo, Tokyo 113-0033, Japan}
\author{Mihiro Asakura}
\affiliation{Department of Physics, The University of Tokyo, Bunkyo, Tokyo 113-0033, Japan}
\author{Shoya Sakamoto}
\affiliation{Institute for Solid State Physics, The University of Tokyo, Kashiwa, Chiba 277-8581, Japan}
\author{Pietro Gambardella}
\affiliation{Department of Materials, ETH Zurich, Honggerbergring 64, 8093 Zurich, Switzerland}
\author{Shinji Miwa}
\affiliation{Institute for Solid State Physics, The University of Tokyo, Kashiwa, Chiba 277-8581, Japan}
\affiliation{Trans-Scale Quantum Science Institute, The University of Tokyo, Bunkyo, Tokyo 113-0033, Japan}
\author{YoshiChika Otani}
\affiliation{RIKEN, Center for Emergent Matter Science (CEMS), Saitama 351-0198, Japan}
\affiliation{Institute for Solid State Physics, The University of Tokyo, Kashiwa, Chiba 277-8581, Japan}
\affiliation{Trans-Scale Quantum Science Institute, The University of Tokyo, Bunkyo, Tokyo 113-0033, Japan}
\author{Satoru Nakatsuji}
\affiliation{Department of Physics, The University of Tokyo, Bunkyo, Tokyo 113-0033, Japan}
\affiliation{Institute for Solid State Physics, The University of Tokyo, Kashiwa, Chiba 277-8581, Japan}
\affiliation{Trans-Scale Quantum Science Institute, The University of Tokyo, Bunkyo, Tokyo 113-0033, Japan}
\author{Christian L. Degen}
\affiliation{Department of Physics, ETH Zurich, Otto-Stern-Weg 1, 8093 Zurich, Switzerland}
\author{Kensuke Kobayashi}
\email{kensuke@phys.s.u-tokyo.ac.jp}
\affiliation{Department of Physics, The University of Tokyo, Bunkyo, Tokyo 113-0033, Japan}
\affiliation{Trans-Scale Quantum Science Institute, The University of Tokyo, Bunkyo, Tokyo 113-0033, Japan}
\affiliation{Institute for Physics of Intelligence, The University of Tokyo, Bunkyo, Tokyo 113-0033, Japan}

\begin{abstract}
Spin chirality in antiferromagnets offers new opportunities for spintronics.
The kagome antiferromagnet \ce{Mn3Sn} is a paradigmatic material in which the antiferromagnetic order parameter can be detected and controlled by electrical means.
However, direct investigation of the magnetic texture of \ce{Mn3Sn} has been challenging because of the tiny moment hosted in its magnetic octupole, hindering further clarification of this unique material.
Here, we address this issue by observing the stray magnetic field from \ce{Mn3Sn} using a diamond quantum scanning magnetometer.
The spatially-resolved intrinsic domains and domain walls in a high-quality single-crystalline \ce{Mn3Sn} film quantitatively reveal the polarization angle of the magnetic octupole in the kagome plane, the domain's local magnetization, the domain wall's width and chirality, {and the octupole order in domain walls.}
Our nanoscale investigation of \ce{Mn3Sn}, a powerful complement to macroscopic measurements, paves the road for developing chiral antiferromagnetism and its potential for spintronic applications.
\end{abstract}

\maketitle

Antiferromagnets (AFs) are promising candidates for realizing high-speed and high-density spintronic devices thanks to their intrinsic terahertz spin dynamics and tiny stray magnetic fields~\cite{JungwirthNatNano2016,BaltzRMP2018}. 
In recent years, much work has been done on the {non-collinear} AF \ce{Mn3Sn} as an ideal platform for AF spintronics~\cite{Nakatsuji2015,Chen2021,Higo2022_rev,Otani2021,Higo2022,Suzuki2017,Higo2018_film,SmejkalNatRevMat2022,Wu2024,rimmler2024}.
\ce{Mn3Sn}, characterized by a cluster magnetic octupole, has vanishingly small net magnetization yet shows a giant anomalous Hall effect comparable to ferromagnetic materials, reflecting the Berry curvature emerging from {the pair of Weyl points in }its topological band structure~\cite{Kuroda2017}. The same feature also leads to an appreciable anomalous Nernst effect~\cite{Ikhlas2017,Li2017}.
The recently reported tunnel magnetoresistance effect~\cite{Chen2023} based on the octupole order\cite{Suzuki2017} further enhances the attraction of this unique material for device applications.

In light of the history of conventional ferromagnet-based spintronics~\cite{Parkin2008, Emori2013}, the importance of a quantitative understanding of the magnetic domains in AFs cannot be overemphasized.
The dynamics of an antiferromagnetic domain are expected to depend on domain wall (DW) chirality (Fig.~1), influencing the manipulation in spintronic devices~\cite{Wornle2021,Emori2013,Li2019}.
\ce{Mn3Sn} film is also promising in this direction\cite{Sugimoto2020}. 
Based on the fundamental research of magnetic octupoles~\cite{Nagamiya1982,Tomiyoshi1982,Tomiyoshi1982a,Brown1990,Kimata2021},   electrical manipulation of the magnetization has been demonstrated using spin-orbit and spin-transfer torque~\cite{Tsai2020,Takeuchi2021,Krishnaswamy2022,Higo2022,Sugimoto2020,Wu2024} and the piezomagnetic effect~\cite{Ikhlas2022_piezo}. 
Obtaining more detailed physical insight into the magnetic domains would accelerate device applications.~\cite{Parkin2008,Nomoto2020,Sugimoto2020,Wu2023}
Particularly for AFs, focusing on the DWs is an effective strategy since they locally break the magnetic symmetry, resulting in an observable stray field.
However, the observation of AF DWs and their chirality are experimentally challenging. 
For \ce{Mn3Sn}, magnetic force microscopy only provides qualitative information~\cite{Liu2023}, while the optical resolution of the magneto-optical Kerr effect (MOKE) measurement is not sufficiently detailed~\cite{Higo2018_moke,Wu2023,Uchimura2022}. 
Recent magnetic field measurements by diamond quantum scanning microscopy (QSM) are promising, as they provide new insights into the DWs of model AF such as \ce{Cr2O3} \cite{Hedrich2021,Wornle2021,Tan2023}; for \ce{Mn3Sn}, despite pioneering observations of polycrystalline samples~\cite{Li2023}, a quantitative investigation has thus far been lacking.

\begin{figure}[tb]
\includegraphics[width=\linewidth]{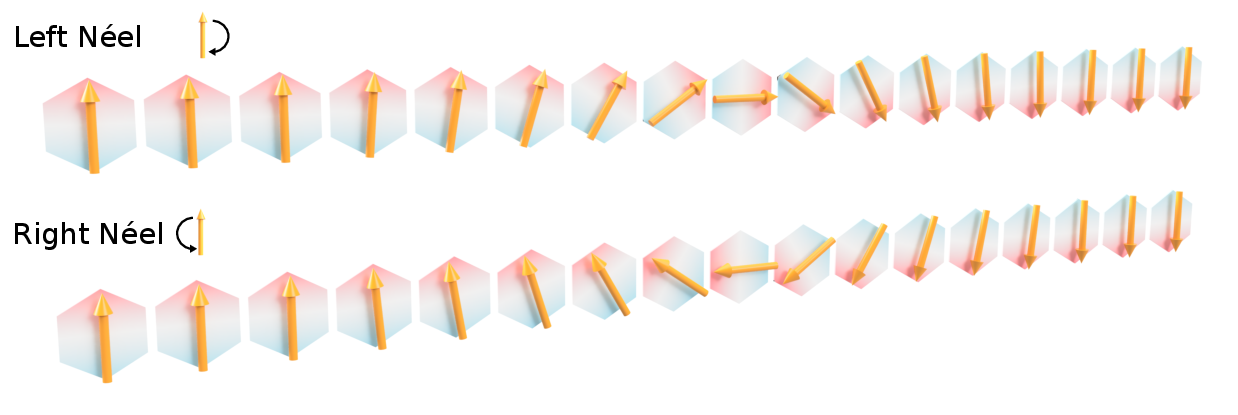}
\caption{Octupole-ordered chiral domain wall.
Schematic illustration of left N\'{e}el wall (top) and right N\'{e}el wall (bottom).
Hexagons indicate the cluster magnetic octupoles, and orange arrows correspond to the octupole moments [also see Fig.~\ref{f1}(b)].
The film structure, shown in Fig.~\ref{f1}(a), breaks the vertical direction symmetry, defining the chirality.
}
\label{f0}
\end{figure}

Here, we report detailed observations of the magnetic domains and chiral DWs in a high-quality single-crystalline \ce{Mn3Sn} film using QSM.
A stray magnetic field from the film directly probes the magnetic octupole ordering in the domains.
The simultaneous detection of the weak magnetization, its axis, and its behavior in the DWs further reveals the nature of the magnetic octupole.
Significantly, the observed DWs, which we determine to be of left N\'{e}el type [top panel of Fig.~\ref{f0}] with a width as narrow as $\sim$40~nm, clearly show how the octupole moment performs a chiral rotation in the kagome plane of \ce{Mn3Sn}.
Our findings will further advance the understanding and control of magnetic octupoles in chiral AFs.

We use a thin \ce{Mn3Sn} film with a thickness $t=20$~nm grown on a 7~nm thick \ce{W} buffer layer on an \ce{MgO} substrate, capped by a 5~nm thick \ce{MgO} layer, as shown in Fig.~2(a).
The films are grown by molecular beam epitaxy.
In \ce{Mn3Sn}, the \ce{Mn} atom forms a kagome lattice and takes an antichiral triangular spin configuration.
As a consequence, the net magnetization is vanishingly small at room temperature~\cite{Tomiyoshi1982a,Brown1990}.
Nevertheless, a slight canting of the spins polarizes a magnetic octupole consisting of six Mn spins [Fig.~2(b)], which results in a detectable magnetic field as if they were magnetic dipoles. 
The ferroic order between them induces a macroscopic magnetization~\cite{Suzuki2017}.
The X-ray diffraction data presented in Figs.~2(c) and 2(d) indicate well-aligned kagome planes.
The MOKE result indicates that the coercive field of the film is $\sim 300$~mT [Fig.~2(e)]. 
Microscopic observation of magnetic domains using MOKE suggests that the crystal grains of the film are several micrometers in size~\cite{Higo2022}.
These results confirm the high crystallinity of the present film.
While the easy axis in the pristine \ce{Mn3Sn} is $\langle 2\bar{1}\bar{1}0\rangle$, the insertion of the W buffer layer increases the perpendicular magnetic anisotropy so that the easy axis in our film is parallel to $[01\bar{1}0]$~\cite{Higo2022}. 

\begin{figure}[tb]
\includegraphics[width=\linewidth]{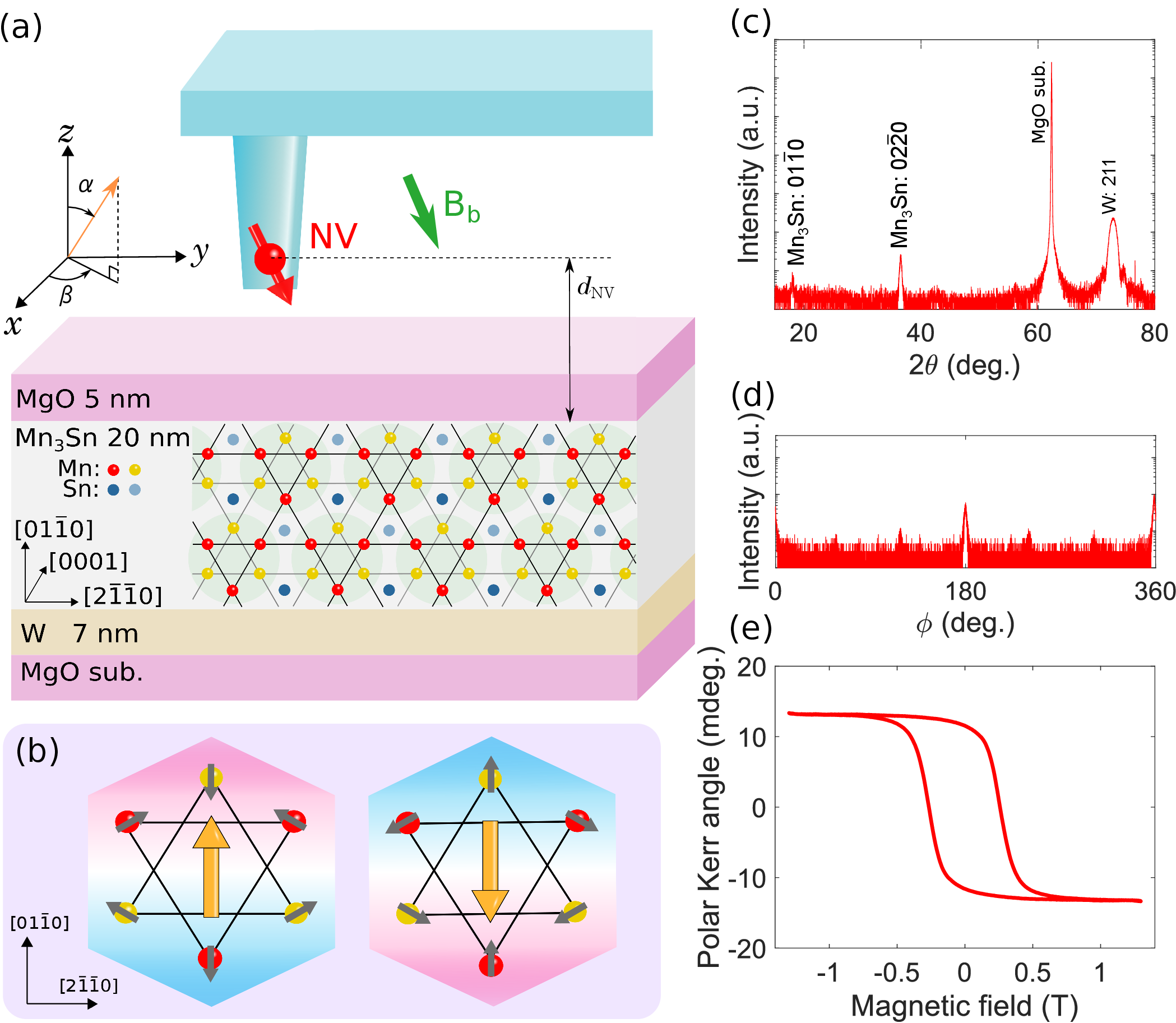}
\caption{\ce{Mn3Sn} thin film and quantum magnetometry.
(a)~Schematic of the sample and QSM tip.
The AB stacking of the crystal lattice is shown inside the \ce{Mn3Sn} film, where the \ce{Mn} atoms in each layer form the kagome lattice.
For QSM, we use a single NV center embedded in the tip, shown by a thick red arrow. 
The bias field $B_b$, shown by a green arrow, is applied along the NV symmetry axis.
The angles $\alpha$ and $\beta$, defined by the orange arrow, are used to analyze the magnetization direction (see text).
(b)~Schematic of the cluster magnetic octupole. 
Each octupole consists of six neighboring \ce{Mn} spins, indicated by the green shading in (a). 
{The gray and orange arrows indicate the dipole moment of \ce{Mn} spin and the octupole moment, respectively.}
The octupole moment direction is upward and downward in the left and right panels, respectively, perpendicularly to the film surface $(01\bar{1}0)$. 
(c) and (d)~XRD $2\theta$-scan and $\phi$-scan data of the \ce{Mn3Sn} thin film, respectively.
(e)~Hysteresis behavior for the z-axis magnetic field measured by MOKE using 660~nm light at room temperature.
}
\label{f1}
\end{figure}

Figure 2(a) also shows a schematic of our QSM (QZabre LLC). The magnetic probe consists of a single nitrogen-vacancy (NV) center embedded in a diamond pillar attached to the tuning fork of the atomic force microscope.
A magnetic field, when present, shifts the discrete energy levels of the NV center by the Zeeman effect.
The shift can be determined by optically detected magnetic resonance using a confocal microscope.
The technique enables us to obtain quantitative and three-dimensional vector information on the magnetic field felt by the NV center~\cite{Wornle2021}.
The NV center scans the $xy$-plane at a height of $d_{\mathrm{NV}}$ [Fig.~2(a)], defined by the distance between the NV center and the top surface of \ce{Mn3Sn} film, yielding the spatial distribution of the magnetic stray field of the sample.
In our work, $d_{\mathrm{NV}}$ is approximately 60~nm, determining the spatial resolution of the imaging (see Sec.~I in SI for details).
For measurement's sake, the external bias field $B_b$ is applied parallel to the NV axis [Fig.~2(a)].
All measurements are performed at room temperature.

\begin{figure*}[tb]
\includegraphics[width=\linewidth]{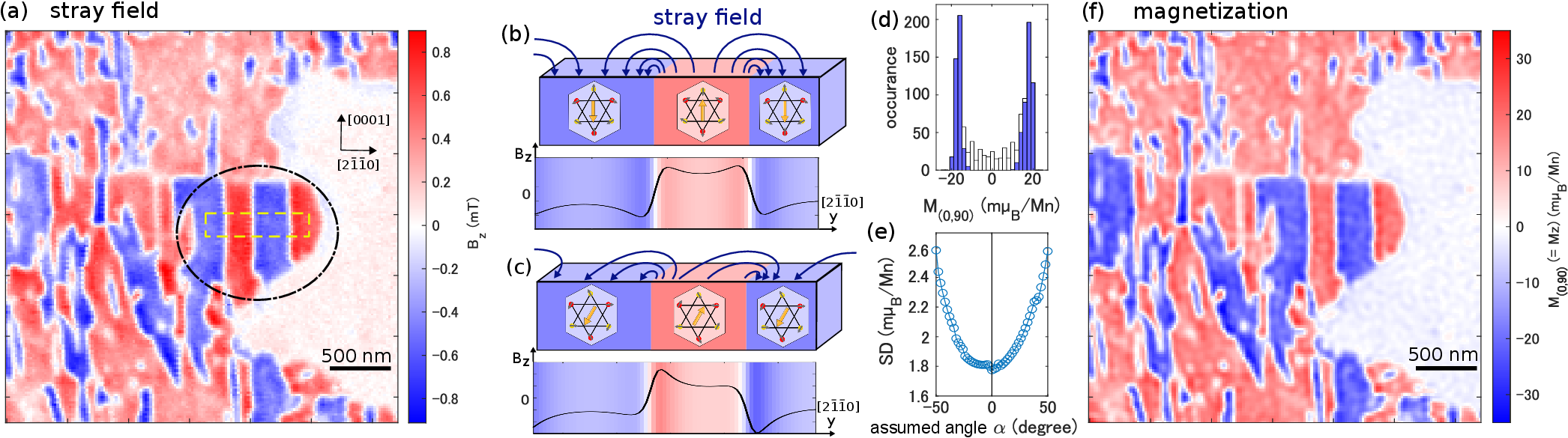}
\caption{Observed stray field and magnetization.
(a)~Image plot of the $z$-axis component of the magnetic stray field $B_z$.
(b) and (c)~Schematics of stray field from a perpendicular magnetization $\parallel [01\bar{1}0]$ and a $30^\circ$~tilted magnetization $\parallel [\bar{1}\bar{1}20]$, respectively. 
The lower panels schematically show the corresponding $B_z$.
The case (b) rather than (c) is realized in the present film.
(d)~Histogram analysis for the reconstructed magnetization in the area surrounded by the dashed yellow line in (a). Blue bars correspond to pixels well inside domains, while white bars are from pixels near the DW's.
(e)~Standard deviation of $|M_{(\alpha,90)}|$ of the blue peaks in (d) as a function of the assumed angle $\alpha$.
(f)~Magnetization image reconstructed into the perpendicular orientation.
}
\label{f2}
\end{figure*}

Figure 3 (a) shows an image plot of the $z$-axis projection of the magnetic field $B_z$, i.e., the stray field perpendicular to the film surface, where the contribution of $B_b$ has been subtracted. 
Most of the image is covered by red and blue regions, corresponding to areas where the field is positive and negative, respectively.
This implies magnetic domain formation due to the perpendicular magnetic anisotropy. 
The field changes steeply at the white lines (zero stray fields) between the red and blue regions, revealing the DWs. 
We also observe  regions with tiny stray fields (white regions) on the right side of the image. 
These may be due to undeveloped magnetic domains or a different magnetization structure from the rest of the image.
From now on, we will focus on the regions with strong magnetic contrast.
The absolute value of the observed stray field from our 20-nm-thick film is only about 0.6~mT.
For comparison, the previous QSM measurement~\cite{Hedrich2020} reported stray fields $\sim 1$~mT for a ferromagnet \ce{CoFeB} film as thin as 1 nm.
Therefore, the present film has a weak magnetization.

Notably, the magnetic domains tend to elongate along the [0001] direction, consistent with a previous MOKE observation~\cite{Uchimura2022}.  
Such a distinctive phenomenon most likely reflects the easy-plane anisotropy of \ce{Mn3Sn}.
In particular, the domains in Fig.~3(a), enclosed by the dash-dot black line, are large and rectangular.
A detailed investigation of them is expected to elucidate the local magnetization texture originating from the magnetic octupole.
In this area, the stray field is almost constant inside the domains, implying that the magnetization is perpendicular to the surface~\cite{Higo2022}; Figure~3(b) schematically shows the stray field for the case with perpendicular magnetic anisotropy, agreeing with the previous observation~\cite{Higo2022}. 
On the other hand, if the magnetization were tilted [Fig.~3(c)], we should observe a field slanting in the domains, which is not the case in Fig.~3(a).

In the case of perpendicular magnetization, a well-known method exists for reconstructing stray fields to magnetization based on the Fourier transform~\cite{Casola2018,Broadway2020}. 
However, for \ce{Mn3Sn}, a more careful analysis without such an assumption is required, as the theory predicts various magnetization directions different from a perpendicular orientation~\cite{Liu2017,Suzuki2017}. 
This analysis will confirm that the magnetization in Fig.~3(a) is predominantly out-of-plane.

First, we expand the existing method~\cite{Casola2018,Broadway2020} to include tilted magnetization cases.
We define ${M_{(\alpha,\beta)}}(x, y)$ as the magnetization value in the film at $(x, y)$ and assume that it is uniaxial with the direction defined by the two angles, $\alpha$ and $\beta$, shown in Fig.~2(a).
Scanning the NV center in the $xy$-plane with $z=d_{\mathrm{NV}}$ fixed [Fig.~2(a)] enables us to obtain the magnetic field amplitude $B_{\mathrm{NV}}(x, y)$ at $(x, y, d_{\mathrm{NV}})$. 
For a sufficiently thin magnetic film, the relation between the magnetization $\mathcal{M}_{(\alpha,\beta)}(k_x, k_y)$ and the magnetic field $\mathcal{B}_{\mathrm{NV}}(k_x, k_y)$ in the corresponding two-dimensional Fourier space is given by
\begin{align}
\mathcal{M}_{(\alpha,\beta)} &= \frac{1}{\tilde{g}(k, d_{\mathrm{NV}},t)}\times \frac{k \mathcal{B}_{\mathrm{NV}}}{ik_x e_x + ik_y e_y - k e_z}  \nonumber \\
&\times\frac{1}{ik_x\sin{\alpha}\cos{\beta} + ik_y\sin{\alpha}\sin{\beta} -k \cos{\alpha}},
\label{eq:reconst}
\end{align}
where $\textbf{\textit{e}}=(e_x, e_y, e_z)$ is the unit vector to define the NV axis [the thick red arrow in Fig.~2(a)], $(k_x, k_y)$ is a wave vector with $k = \sqrt{k_x^2 + k_y^2}$, and $\tilde{g}$ is a Green's function in Fourier space. 
The measured $B_{\mathrm{NV}}$ is converted to $\mathcal{B}_{\mathrm{NV}}$, then to $\mathcal{M}_{(\alpha,\beta)}$ using Eq.~(\ref{eq:reconst}), and finally to ${M_{(\alpha,\beta)}}$ (see Secs.~II and III in SI for detail).

Next, we perform a histogram analysis of the statistical distribution of the obtained magnetization~\cite{Sun2021,Appel2019} to search for the most likely magnetization direction.
Figure~3(d) shows the histogram obtained for the magnetization surrounded by the dashed yellow line in Fig.~3(a).
Most data are clustered at either the positive or negative extremum, forming two sharp peaks with the same absolute value (blue bars). 
Meanwhile, values in-between reflect areas of low perpendicular magnetization near DWs (white). 
These observations imply that the domains in this area have antiparallel octupole moments of the same strength.
Sharper peaks correspond to better binarization and, thus, a more likely case. 
Figure~3(e) plots the standard deviation of $|{M_{(\alpha,90)}}|$ obtained by varying $\alpha$ from $-50^{\circ}$ to $50^{\circ}$ with $\beta=90^{\circ}$ fixed for the same area.
The fact that it is minimal at $\alpha = 0^{\circ}$ indicates that a perpendicular magnetization is the most likely.
We have repeated the same analysis assuming various possible combinations of $\alpha$ and $\beta$ and have confirmed that $\alpha=0^{\circ}$ and $\beta=90^{\circ}$ give the smallest deviation.
The resulting magnetization configuration is shown in Fig.~3(f).
Our observations independently confirm~\cite{Higo2022} that inserting a W layer favors perpendicular magnetic anisotropy.

The magnetization obtained by averaging $|{M_{(0,90)}}|$ excluding DWs regions [Fig.~3(e)] is $17\pm2~\mathrm{m\mu_B/Mn}$.
This value is two orders of magnitude smaller than the typical magnetization value of $1$--$3~\mathrm{\mu_B/Mn}$ in a Mn-based ferromagnet.
Our value is about three times larger than the $\sim 6~\mathrm{m\mu_B/Mn}$ reported for a \ce{Mn3Sn} film by a standard magnetization measurement~\cite{Higo2022}.
The difference is reasonable, as our QSM is performed on well-developed magnetic domains and not averaged over the entire film. 
Rather, our measurements demonstrate that the intrinsic nature of the material can be revealed by measuring high-quality portions with a precise local probe.

Systematic investigation of the response of \ce{Mn3Sn} to magnetic fields elucidates the nature of magnetic domains and DWs.
Figures 4(a), 4(b), and 4(c) show the $z$-projection of the stray field $B_z$ when the bias magnetic field is sequentially varied from 9.5~mT to 26.5~mT to 10.2 mT, respectively. 100 mT is applied once between Figs.~4(b) and 4(c). 
The bias field is subtracted in these images. 
The line scans at the black dashed lines in Figs.~4(a), 4(b), and 4(c) are shown in Fig.~4(d).
As the bias field increases from Fig.~4(a) to Fig.~4(b), the blue (red) domains with the magnetization oriented parallel (anti-parallel) to the field expand (shrink).
The corresponding two DWs, DW1 and DW2, shift tens of nm along opposite $[2\bar{1}\bar{1}0]$ directions, shown in Fig.~4(d).
The fact that the DWs already move at $B_b = 26.5~\mathrm{mT}$, a field much smaller than the coercivity of 300~mT [Fig.~2(e)], indicates that the DW dynamics are mainly governed by exchange interaction and magnetic anisotropy rather than by pinning potentials due to grain boundaries.

\begin{figure*}[tb]
\includegraphics[width=1.0\linewidth]{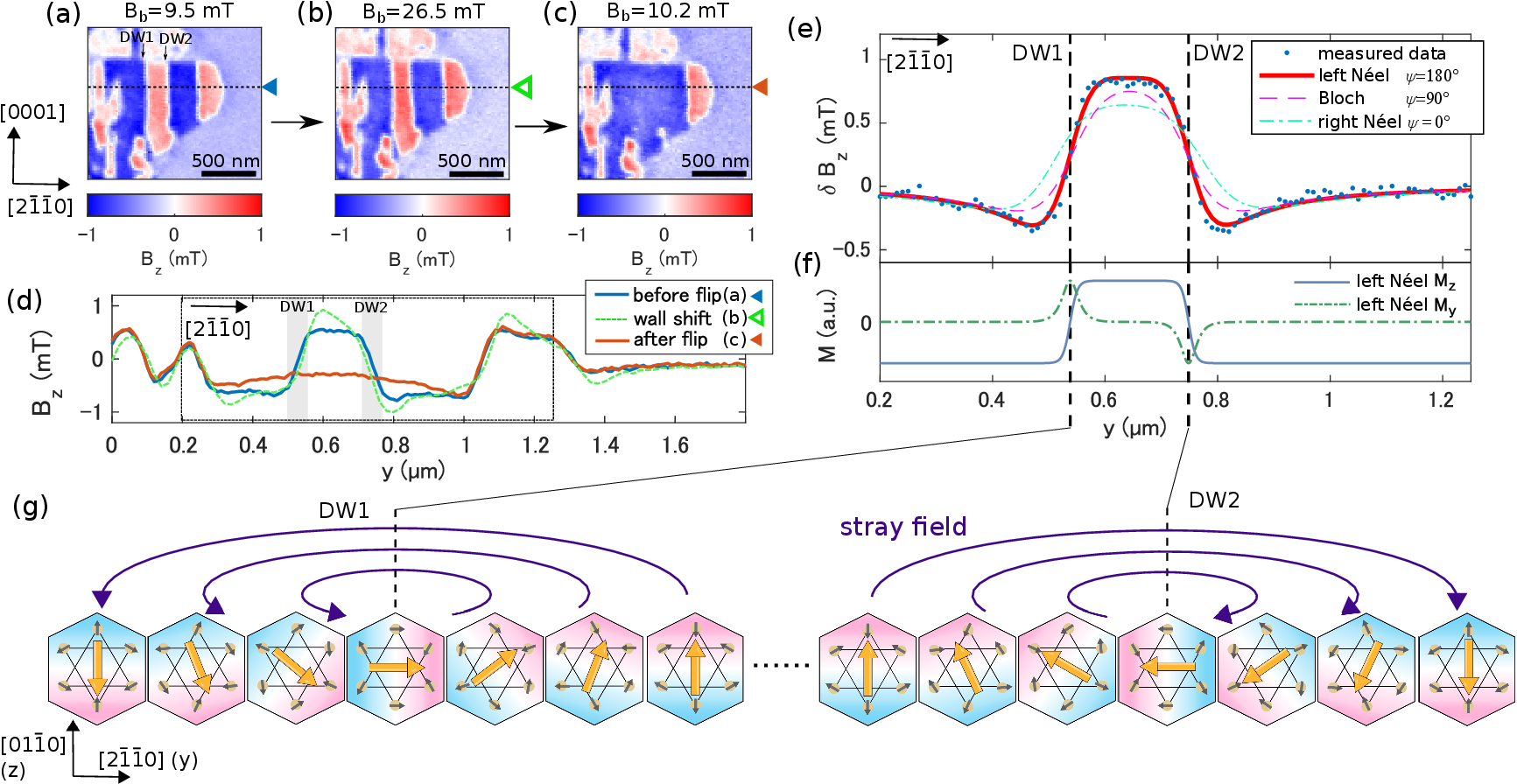}
\caption{Observation of chiral domain wall.
(a-c)~$z$-projection of the stray field when the bias magnetic field is sequentially varied from 9.5~mT to 26.5~mT to 10.2 mT, respectively. 100 mT is applied once between (b) and (c).
(d)~Blue, green, and orange lines are cross-sections of the dashed black lines in (a), (b), and (c), respectively.
DW1 and DW2 are indicated in a gray shade.
(e)~Shape of DW1 and DW2 extracted as the magnetic field difference between (a) and (c) is shown as blue dots. 
The fitted lines of left N\'{e}el ($\psi_i = 180^{\circ}$), Bloch ($\psi_i = 90^{\circ}$), and right N\'{e}el ($\psi_i = 0^{\circ}$) walls are drawn by red, magenta, and cyan lines, respectively.
(f)~$y$- and $z$- projections of the magnetization of the left N\'{e}el fitting. 
Vertical black dashed lines indicate the positions of DW1 and DW2.
(g)~Schematics of magnetization and stray field lines near DW1 (left) and DW2 (right). The schematic of DW2 is equivalent to the top panel of Fig.~1.
}
\label{f5}
\end{figure*}

The red domain at the center vanishes in Fig.~4(c) after 100~mT is applied.
The magnetization of this domain, initially surrounded by DW1 and DW2, flips in the same direction as the neighboring domains.
The flipping field of the domain is at only one-third of the coercive field [measured by MOKE, Fig.~2(e)], again pointing toward on the intrinsic nature of the observed domains.
The obtained field profile shown in Fig.~4(d) clarifies that the magnetic field far from this region is almost unchanged between Figs.~4(a) and 4(c), confirming the high reproducibility and the noninvasiveness of the QSM measurement. 
Thus, we take the difference between the two for further analysis, as shown in Fig.~4(e) with blue dots.
The resultant field $\delta B_z$ originates purely from DW1 and DW2, corresponding to the octupole moment direction in the DWs.
We notice that each wall width is significantly smaller than 200~nm, clearly much smaller than the reported wall width of $\sim 630$~nm~\cite{Sugimoto2020,Higo2022}.

We analyze the shape of the two DWs by fitting them using the conventional DW model given by~\cite{Tetienne2015,Wornle2021,Velez2019}
\begin{align}
{M_x} =&   -\frac{M_s\sin{\psi}}{\cosh\left(\frac{y-y_1}{\Delta_{\mathrm{1}}}\right)} + \frac{M_s\sin{\psi}}{\cosh\left(\frac{y-y_2}{\Delta_{\mathrm{2}}}\right)}. \label{eq:mx}\\
{M_y} =&   -\frac{M_s\cos{\psi}}{\cosh\left(\frac{y-y_1}{\Delta_{\mathrm{1}}}\right)} + \frac{M_s\cos{\psi}}{\cosh\left(\frac{y-y_2}{\Delta_{\mathrm{2}}}\right)}. \label{eq:my}\\
{M_z} =&  M_s \tanh\left(\frac{y-y_1}{\Delta_{\mathrm{1}}}\right) - M_s \tanh\left(\frac{y-y_2}{\Delta_{\mathrm{2}}}\right) - M_s, \label{eq:mz}
\end{align}
The fitting parameters are the positions $y_i$ and widths $\pi\Delta_i$ of DW$i$ ($i=1, 2$), with the saturation magnetization $M_s$ being common for DW1 and DW2.
While the magnetization away from the DWs is parallel to the $z$-direction, it has a finite $xy$-plane component characterized by the angle $\psi$ inside DWs. 
$\psi=0^{\circ}$ corresponds to the positive direction of the $y$-axis.
The stray field from the DWs is determined by $M_y$ and $M_z$ due to the translational symmetry along the $x$-axis $\parallel$ [0001] direction.
Fitting the DW model to the magnetic field data with $\psi = 180^{\circ}$, as shown in red in Fig.~4(e), we obtain $M_s=6.9\pm0.3~\mathrm{kA/m} = 16.1\pm0.2~\mathrm{m\mu_B/Mn}$, $\Delta_1 = 13\pm3$~nm, and $\Delta_2=12\pm3$~nm.
The obtained $M_s$ is consistent with the $17\pm2~\mathrm{m\mu_B/Mn}$ discussed above in Fig.~3.
The fittings with $\psi = 90^{\circ}$ and $0^{\circ}$ shown in magenta and cyan lines in Fig.~4(e), respectively, are unsuccessful.

The observation of $\psi = 180^\circ$ tells us that the DWs are the left N\'{e}el type, rather than the right N\'{e}el type ($\psi = 0^\circ$) or the Bloch type ($\psi = 90^\circ$).
In the N\'{e}el cases ($\psi = 0^\circ$ or $180^\circ$), $M_x=0$ according to Eq.~(\ref{eq:mx}), and the octupole moments with constant magnitude rotate in the kagome plane around the $x$-axis.
In the left N\'{e}el wall case, the moment at the DW center is horizontal and faces left when the upward magnetization region is placed on the left side~(the top panel of Fig.~1).
The spatial variations of the magnetization $M_y$ and $M_z$ are shown in Fig.~4(f).
$M_z$ changes its sign at the center of DWs, while $M_y$ is finite only inside the DWs.
There are several {other} theoretically possible DW structures~\cite{Wu2023,Malozemoff2016}. 
We have performed fittings assuming that the magnetization can vary inside the DW and that the DWs split into several parts.
The fittings confirmed that the left N\'{e}el type is the most likely to explain the observed magnetic field (see Sec.~X in SI for details).
Figure~4(g) schematically depicts the structure of the DWs.
{The octupole ordering in DWs suggests that the spins at each site in the octupole also have a ferroic order.}
While the magnetic field distribution depicted in Fig.~4(d) contains several DWs, rigorous analysis is possible only for DW1 and DW2 due to the interference of the fields from multiple DWs. 
Nevertheless, the observed distribution over 17 DWs can be quantitatively reproduced by assuming the left N\'{e}el-type DWs with the fitted parameters (see Sec.~XIII in SI for details). 
Our data, therefore, suggest that the DWs along [0001] in this {\ce{Mn3Sn}/W} film are mainly of the left N\'{e}el-type type.

Theoretically, the DW width $\pi\Delta = \pi\sqrt{A/\kappa} $ along [0001] of \ce{Mn3Sn} is determined by minimizing the energy density~\cite{Malozemoff2016},
\begin{align}
\kappa =K + \frac{1}{2} \mu_0 M_s^2 \cos^2\psi + K_p\sin^2\psi
\label{eq:kappa}
\end{align}
where $A = 0.568~\mathrm{meV/\AA}$ is the stiffness constant~\cite{Liu2017}, $K = 1.4\times10^{-7}~\mathrm{meV/\AA^3}$ is the magnetic anisotropy~\cite{Higo2022}, $K_p$ is the in-plane anisotropy, $\psi$ is the DW angle, and $\mu_0$ is the vacuum permeability.
The second term of Eq.~(\ref{eq:kappa}) is the magnetostatic energy under zero magnetic field.
The effect of bias magnetic field on $\Delta$ is neglected because $\Delta$ changes by at most 1~nm under the magnetic fields used in this study, which is smaller than the error bar of the fitting.
We use $M_s = 6.9\pm0.3~\mathrm{kA/m}$ obtained above.
The left N\'{e}el wall corresponds to $\psi = 180^{\circ}$. 
Thus, in Eq.~(\ref{eq:kappa}), the third term is zero, while the second term is finite.
This magnetostatic energy, in turn, is a hundred times larger than the magnetic anisotropy $K$, meaning that it determines the wall width. 
The DW width is predicted theoretically as $\Delta =17\pm 2$~nm, a value very close to the experimental result of Fig.~4(e). 
Considering that the calculation uses the known values except for $M_s$ and $\psi$, the consistency validates the present result and analysis using the conventional DW model to \ce{Mn3Sn}.
Further discussion, including the demagnetizing factor, suggests that the magnetostatic energy of a N\'{e}el DW is smaller than that of a Bloch wall~\cite{Neel1955}.
This energy difference leads to a tendency for the domains to be separated by DWs running along the [0001] direction, resulting in the appearance of elongated domains as shown in Fig.~3(a) and the previous study~\cite{Uchimura2022} (see Sec.~XII in SI for detail).

We address the implication of the present experimental findings regarding AF spintronics applications of \ce{Mn3Sn}.
The observation of the domains by QSM proves the existence of a perfectly perpendicular magnetization area, which maximizes the anomalous Hall effect and reduces the power consumption for readout.
The observed magnetization of the well-developed domains is greater than previous values~\cite{Sugimoto2020,Chen2023}, promising accelerated writing and reading speed of devices.
The  coercive field, which is locally smaller than the area-averaged MOKE result, suggests a smaller threshold of intrinsic magnetization reversal than previous reports, enabling high-efficiency spin-orbit torque switching~\cite{Higo2022}.
The identified left N\'{e}el wall chirality is crucial for developing DW devices, as the direction of current-induced DW motion depends on the chirality.
The development of chirality may be due to the symmetry breaking in the vertical direction by inserting a W layer.
Replacing W with other heavy metals, such as platinum, might invert the chirality and the DW motion direction.
Considering that the DW width ultimately determines the device size, the narrow width of $\pi\Delta\sim 37$~nm obtained in our study redefines the integration limit and scalability of \ce{Mn3Sn} devices.
{These parameters can contribute to revealing the swapping trajectory of Weyl points at the DW, which was theoretically discussed recently.}\cite{Araki2020,Ozawa2024}
Finally, {the octupole ordering in DWs is the basis for clarifying the detailed mechanism of switching the octupole moment using spin transfer and spin-orbit torque.}

To conclude, our QSM measurement experimentally established the weak magnetization, the direction of the octupole moment, and the chiral behavior of the moment inside the kagome plane of \ce{Mn3Sn}, all vividly illustrating the role of its cluster magnetic octupoles.
While the weak magnetization of such thin films is challenging to measure, accurate values are obtained by the local method of QSM.
The magnetic parameters provide quantitative input for the AF spintronics applications based on this material.
The experimental proof of chiral DWs is a powerful complement to macroscopic measurements and further advances our understanding and control of chiral antiferromagnetism.

\begin{acknowledgments} 
We thank S.~Fukami, C.~Broholm, Y.~Araki, and A.~Ozawa for useful discussions.
The experiment was supported by QZabre LLC. 
This work was partially supported by Grants-in-Aid for Scientific Research (Nos.~JP23K25800, JP22K03524, JP22KJ1059, and JP22J21412); Swiss National Science Foundation, Grant Nos. 200020\_200465 and 200020\_212051/1; Japan Science and Technology Agency, JST-CREST (JPMJCR23I2), JST-MIRAI Program (PMJMI20A1), JST-ASPIRE (JPMJAP2317), Japan; Daikin Industries, Ltd.; Kondo Memorial Foundation; the Cooperative Research Project of RIEC, Tohoku University. M.T. acknowledges financial support from World-leading Innovative Graduate Study Program for Forefront Physics and Mathematics Program to Drive Transformation, The University of Tokyo. Z.X. acknowledges MSCA Project SPEAR, Grant No. 955671, funded by the European Commission through H2020.
\end{acknowledgments} 

\bibliography{Mn3Sn.bib}

\end{document}